\documentclass[10pt]{article}

\usepackage{amsmath,amssymb,dsfont,slashed,color,hyperref}
\usepackage{authblk}
\usepackage[toc,page]{appendix}
\usepackage{graphicx,caption,subcaption,tikz}

\usepackage{geometry} 
\usepackage{tikz} 
\usepackage{lipsum}

\newlength\imagewidth
\newlength\imagescale

\newcommand{\half}{\frac{1}{2}}

\newcommand{\JJ}{\mathbb{J}}
\newcommand{\PP}{\mathbb{P}}

\newcommand{\TT}{\mathbb{T}}
\newcommand{\ZZ}{\mathbb{Z}}

\newcommand{\QQ}{\mathbb{Q}}
\newcommand{\QQb}{\overline{\mathbb{Q}}}

\newcommand{\AAA}{\mathbb{A}}

\newcommand{\se}{\slashed{e}}

\newcommand{\lD}{\overleftarrow{D}}

\newcommand{\psibar}{{\overline{\psi}}}
\newcommand{\chibar}{{\overline{\chi}}}

\newcommand{\epsee}{\epsilon^a{}_{bc} e^b e^c}

\newcommand{\rlfrac}{\left(\frac{r}{\ell}\right)}

\newcommand{\cecs}{Centro de Estudios Cient\'ificos, Arturo Prat 514, Valdivia, Chile}
\newcommand{\uss}{Universidad San Sebasti\'an, General Lagos 1163, Valdivia, Chile}
\newcommand{\uantof}{Departamento de F\'isica, Universidad de Antofagasta, Aptdo. 02800, Chile}

\begin{document}

\title{Gravitating spinor torsional BTZ solution}

\author[1,2]{P. D. Alvarez \thanks{E-mail: \href{mailto:pedro.alvarez@uss.cl}{\nolinkurl{pedro.alvarez@uss.cl}}}}
\author[3]{J. Ortiz \thanks{E-mail: \href{mailto:juan.ortiz@uantof.cl}{\nolinkurl{juan.ortiz@uantof.cl}}}}
\author[1,2]{J. Zanelli \thanks{E-mail: \href{mailto:z@cecs.cl}{\nolinkurl{z@cecs.cl}}}}

\affil[1]{\uss}
\affil[2]{\cecs}
\affil[3]{\uantof}

\date{}
\maketitle

%\date{\today}

\begin{abstract}

We construct exact analytic solutions for a supersymmetric Chern–Simons theory with matter fields in the adjoint representation (unconventional supersymmetry). The configurations correspond to gravitating spinor fields on BTZ geometries with torsion. Although the spinor profiles exhibit divergent behavior near the black hole horizon, they yield finite contributions to conserved charges. We demonstrate that the spinors solve the full nonlinear field equations and satisfy integrability conditions derived from the Bianchi identities. The resulting energy–momentum tensor is traceless, consistent with the Weyl invariance of the theory. We also analyze spinor solutions based on $adS_3$-Killing spinors and identify the conditions under which such backgrounds can support nontrivial spinor configurations. Our findings open the way to new fermionic sectors in 3D gravity with torsion, distinct from previously known perturbative deformations.
 
\end{abstract}

\maketitle

%%%%%%%%%%%%%%%%%%%%%%%%%%%%%%%%%%%%%%%
\section{Introduction}\label{int}
%%%%%%%%%%%%%%%%%%%%%%%%%%%%%%%%%%%%%%%

In this paper we examine the possibility of having exact solutions of a three-dimensional Chern-Simons model for unconventional supersymmetry (USUSY). In this model, the matter fields are in the adjoint representation \cite{Alvarez:2011gd,Alvarez:2015bva} and matching between the degrees of freedom of bosons and fermions \cite{Sohnius:1985qm} is not required.

Three-dimensional theories of gravity admit a variety of black hole solutions incorporating matter which might be viewed as toy models that offer insights into various astrophysical and cosmological scenarios. These include polytropic stars as studied in \cite{Sa:1999yf}, rigidly rotating perfect fluid stars \cite{Gundlach:2020ovt}, and low-mass strange stars modeled using the Heintzmann ansatz \cite{Murshid:2021iaq}. Other noteworthy examples are circular thin shells in 2+1-dimensional $F(R)$ gravity \cite{Eiroa:2020dip}, as well as exact solutions in (2+1)-dimensional anti-de Sitter space-time with linear or nonlinear equations of state (e.g., \cite{Banerjee:2014mwa} and references therein). In \cite{Alvarez:2022bqe}, a gravitating spinor solution with torsion that drives an exponential expansion was found. 

The USUSY construction incorporates spin-1/2 matter fields as part of a gauge connection in the adjoint representation, associated to the generators of supersymmetry transformations,
\begin{equation}
 \QQb\se\psi+\psibar\se\QQ \subset \mathbb{A}\label{connectionintro}
\end{equation}
where $\se =e^a\gamma_a$ and $e^a = e^a{}_\mu dx^\mu$ are the local orthonormal frame 1-forms (vielbein) and $\psibar \equiv i \psi^\dagger \gamma^0$ is the Dirac adjoint. The composite field $e^a \gamma_a \psi$ is a fermionic one-form and is annihilated by the spin-3/2 projector \cite{Alvarez:2011gd,Alvarez:2013tga,Alvarez:2015bva,Alvarez:2020qmy,Alvarez:2020izs,Alvarez:2021zhh,Alvarez:2021zsw}. Unconventional supersymmetry actions based on the Chern-Simons forms can be defined in all odd dimensions. In \cite{Alvarez:2014uda} and more recently in \cite{Andrianopoli:2024twc}, geometries for the three dimensional black hole with torsion were discussed (see also \cite{Alvarez:2015bva} for nontrivial $SU(2)$ solutions). 

The question we would like to address here is whether there exist nontrivial exact fermionic solutions in the three-dimensional spacetime described by the Einstein equations. We will be interested in the simplest nontrivial geometry and in the absence of other gauge fields.  The geometry is assumed to be of the BTZ family: circularly symmetric and stationary, characterized by two real integration constants, the mass $M$ and the angular momentum $J$. If regular solutions of the Dirac equation exist that can be switched on and off for all values of $M \geq J$, they would correspond to a spin-1/2 hair on the $2+1$ black hole. Our approach here is nonperturbative and can therefore be seen as complementary to other methods in the literature like \cite{Dasgupta:1998jg,Gentile:2013nha}.

The paper is organized as follows. In sect.~\ref{sec:1}, we define the basics of the model. In sect. \ref{fieldeqs}, we present the field equations and we study integrability conditions on the spinorial bilinears coming from the on-shell restriction of the Bianchi identities. In sec. \ref{spinoreqs}, we solve the spinor equations and compute the charges of the theory. In sect.~\ref{sec:conclu}, we summarize our results and provide some ideas for future developments.

%%%%%%%%%%%%%%%%%%%%%%%%%%%%%%%%%%%%%%%%%%%%%%%%%%%%%%%%%%%%
\section{USUSY: SUSY in the adjoint representation}\label{sec:1}
%%%%%%%%%%%%%%%%%%%%%%%%%%%%%%%%%%%%%%%%%%%%%%%%%%%%%%%%%%%%

The supersymmetric extension of the Lorentz group in $2+1$ dimensions can be constructed enlarging the Lorentz algebra with the inclusion of some supersymmetry generators $\QQ^\alpha_i$, where $\alpha=1,2$ is a spinor index and $i=1,2,...,n$ is an internal group. In this way, $\QQ$ is in a spin-1/2 representation of the Lorentz group and in the fundamental (vector) representation of the internal group. If $\JJ_{ab}$ and $\TT_I$ are the Lorentz and internal symmetry generators, we expect
\begin{align}
[\JJ_{ab}, \QQ^\alpha_i]\,&= \frac{1}{2}(\gamma_{ab})^\alpha_{\;\;\beta} \; \QQ^\beta_i\,,  \qquad \;\; [\JJ_{ab}, \QQb_\alpha^i] \,= \frac{1}{2}(\gamma_{ab})^\beta{}_\alpha \; \QQb_\beta^i\,,  \\
[\TT_I,\QQ^\alpha_i]\,&= (\sigma_I)_i^{\;\;j}\; \QQ^\alpha_j\,, \qquad\qquad [\TT_I, \QQb_\alpha^{\;\;i}]\,= -(\sigma_I)_j{}^i\; \QQb_\alpha^j \,,
\end{align}
where $\gamma_{ab}=\frac{1}{2}[\gamma_a, \gamma_b]= \epsilon_{ab}^{\ \ c}\gamma_c$ The Lorentz and internal generators, on the other hand, satisfy their respective Lie algebras, $[\JJ_{ab}, \JJ_{cd}]\,=\, f_{ab\, cd}{}^{ef} \;\JJ_{ef}$, $[\TT_I, \TT_J ] \,=\, f_{IJ}{}^L \TT_L$. The crucial point is the closure of the superalgebra, which requires the anti-commutator of the supersymmetric charges to close on the bosonic generators,
\begin{equation}
\{\QQb_\alpha^i, \QQ^\beta_j\} \propto  \delta^i_{\;j}\,\left(\gamma^{ab}\right)^\beta_{\;\,\alpha}\, \JJ_{ab} + \delta^\beta_{\; \alpha} (\sigma^I)^i_{\;j} \TT_I\,.
\end{equation}
From this algebra, a (super) connection is defined as 
\begin{equation}
\AAA \sim \frac{1}{2}\omega^{ab} \JJ_{ab} + A^I \TT_I + \QQb_\alpha^i \xi^\alpha_i + \overline{\xi}_\alpha^i\,\QQ^\alpha_i\,.
\end{equation}
The conventional procedure would be to assume $\xi$ a spinorial one-form (gravitino), while in USUSY one assumes $\xi$ to be the product of a vielbein one-form $e^a=e^a{}_\mu dx^\mu$, a Dirac gamma matrix $\gamma_a$ and a spin-1/2 zero-form $\psi$ representing matter fields similar to leptons and quarks,
\begin{equation}
    \xi\,=\ dx^\mu \xi_\mu\,= \, dx^\mu\, e^a_\mu \gamma_a \psi\,,
\end{equation}
(here the spinor indices have been suppressed). Finally, with this connection a Yang-Mills action can be produced in any dimension, but in odd dimensions there is the additional possibility of defining  a Chern-Simons action. Here we will consider a model in 2+1 dimensions with an internal symmetry given by two copies of $SU(2)\times U(1)$ that will be labeled by an index $+$ or $-$ (see Appendix \ref{apprep} for details).

Let us consider a model with matter in the adjoint representation for an $su(2|2)\times su(2|2)$ gauge potential
\begin{equation}
 \AAA=\AAA_+ + \AAA_-\,,
\end{equation}
with
\begin{align}
 \AAA_+&=\half \omega^{ab}_+ \JJ^+_{ab}+A^I_+ \TT^+_I+\QQb^{+i}\se\psi_i+\overline{\psi}^i\se\QQ^+_i+b_+ \ZZ^+\,,\label{Aplus}\\
 \AAA_-&=\half \omega^{ab}_- \JJ^-_{ab}+A^I_- \TT^-_I+\QQb^{-i}\se\chi_i+\overline{\chi}^i\se\QQ^-_i+b_- \ZZ^-\,.\label{Aminus}
\end{align} 
With this connection an Ach\'ucarro-Townsend--like action can be defined as \cite{Achucarro:1986uwr},
\begin{equation}\label{action}
 S=2\ell\int \left(L_\text{CS}(\AAA_+)-L_\text{CS}(\AAA_-)\right)\,.
\end{equation}
where the Chern-Simons Lagrangian is
\begin{equation} \label{CS-Lag}
L_\text{CS}(\AAA)=\frac{\kappa}{2}\langle \AAA d\AAA+\frac{2}{3}\AAA^3\rangle\,.
\end{equation}
and the brackets $\langle...\rangle$ represent the supertrace in the conformal superalgebra representation provided in Appendix \ref{apprep}. The canonical length-dimensions of the fields in $d=3$ are: $[A^I_\mu]=l^{-1}$, $[f^a_\mu]=l^0=[e^a_\mu]$, $[\psi]=l^{-1}$.

In three dimensions, the Einstein-Hilbert action plus a cosmological constant term, with $\Lambda = -1/\ell^2$, is recovered for
\begin{align}
 \omega_\pm^{ab} & = \omega^{ab} \mp \frac{1}{\ell}\epsilon^{ab}{}_c f^c\,,\label{wpm}\\
 \JJ^\pm_{ab} & = \frac{1}{2}(\JJ_{ab} \mp \ell \epsilon_{ab}{}^c \PP_c)\,.
\end{align}
In analogy with (\ref{wpm}), we take
\begin{equation}
 A^I_\pm=A^I\pm B^I\,,
\end{equation}
which implies that $B^I=\frac{1}{2}(A^I_+-A^I_-)$ transform as a one-form under the diagonal $SU(2)$ gauge symmetry. The action (\ref{action}) takes the form
\begin{equation}
 S=2\ell\int \left(L_\text{CS}(\omega_+)-L_\text{CS}(\omega_-)+L_\text{CS}(A_+)-L_\text{CS}(A_-)+L_\psi^+-L_\chi^-\right)\,,\label{fullaction}
\end{equation}
where
\begin{align}
L_\text{CS}(\omega_+)-L_\text{CS}(\omega_-) =& \frac{\kappa}{\ell}(\half  \epsilon_{abc}f^a R^{bc}+\frac{1}{6\ell^2}\epsilon_{abc}f^a f^b f^c+\frac{1}{2}d(f^a \omega_a))\,,\\
L_\text{CS}(A_+)-L_\text{CS}(A_-) =& \kappa(B^I F_I+\frac{1}{6}\epsilon_{IJK}B^I B^J B^K+\frac{1}{2}d(B^I A_I))\,,\\
L_\psi^+ - L_\chi^- =& \frac{\kappa}{2}(\psibar\se )(\overleftarrow{D}^+ - D^+)(\se\psi) - \frac{\kappa}{2}(\chibar\se )(\overleftarrow{D}^- - D^-)(\se\chi)\,.
\end{align}
The spinor lagrangian can be further expanded in terms of the Lorentz covariant derivatives using
$L_\psi^\pm=L_\psi+\Lambda^\pm_\psi+\Gamma^\pm_\psi$, where
\begin{align}
 L_\psi&=\frac{\kappa}{2}(\overline{\psi}\se )(\overleftarrow{D}-D)(\se\psi)=\frac{\kappa}{2}\overline{\psi}(\se\se D-\overleftarrow{D}\se\se )\psi-\frac{\kappa}{2}\overline{\psi}(\se\slashed{T}+\slashed{T}\se )\psi\,,\\
 \Lambda^\pm_\psi&=\mp\frac{\kappa}{2\ell}\overline{\psi}\se\slashed{f}\se\psi\,,\\
 \Gamma^\pm_\psi&=\pm i\frac{\kappa}{2}\overline{\psi}\se B^I\sigma_I \se\psi\,.
\end{align}
The $so(1,2)\times su(2)$ covariant derivative in the spin-$1/2$ representation\footnote{If $\Omega^m$ is an $m$-form, then $\Omega^m\overleftarrow{d}=(-1)^m d\Omega^m$.}
\begin{eqnarray}\label{D}
 D_i^{\ j}&=&\delta^j_i d+\frac{1}{2}\delta^j_i\omega^{ab}\Sigma_{ab}-\frac{i}{2}A^I(\sigma_I)_i^{\ j}\,,\\
 \overleftarrow{D}_i^{\ j}&=&\overleftarrow{d} \delta^j_i-\frac{1}{2}\omega^{ab}\Sigma_{ab} \delta^j_i +\frac{i}{2}A^I(\sigma_I)_i^{\ j}\,.
\end{eqnarray}
This covariant derivative acts along the diagonal component of two copies of $so(1,2)\times su(2)$. The full Lagrangian for the spinors is
\begin{align}
L_\psi^+-L_\chi^-= &\kappa\left[\frac{1}{2}\psibar(\epsilon_{abc}e^a e^b \gamma^c D-\lD \epsilon_{abc}e^a e^b \gamma^c)\psi- e^a T_a\psibar\psi \right.\nonumber \\
&-\frac{1}{2\ell}\epsilon_{abc}e^ae^bf^c\psibar\psi-\frac{i}{2}\epsilon_{abc}e^ae^bB^I \psibar \gamma^c\sigma_I\psi\nonumber \\
&-\frac{1}{2}\chibar(\epsilon_{abc}e^a e^b \gamma^cD-\overleftarrow{D}\epsilon_{abc}e^a e^b \gamma^c)\chi+ e^a T_a\chibar\chi \nonumber\\
&\left. -\frac{1}{2\ell}\epsilon_{abc}e^ae^bf^c\chibar\chi-\frac{i}{2}\epsilon_{abc}e^ae^bB^I\chibar\gamma^c\sigma_I\chi\right]\,. \label{spinorlag}
\end{align}

%%%%%%%%%%%%%%%%%%%%%%%%%%%%%%%%%%%%%%%%%%
\section{Field equations}\label{fieldeqs}
%%%%%%%%%%%%%%%%%%%%%%%%%%%%%%%%%%%%%%%%%%

In this section we will discuss generic exact solutions of the field equations.

Varying (\ref{fullaction}) with respect to $f^a$ and $\omega^{ab}$ gives
\begin{align}
 &R^{ab}+\frac{1}{\ell^2}f^af^b - \Phi e^a e^b = 0\,,\label{eomf}\\
 &\frac{1}{\ell}T_f^a - \frac{1}{2}\Delta\epsilon^a_{\ bc}e^be^c = 0\,,\label{eomw}
 \end{align}
where
\begin{align}
 \Delta =& \psibar \psi - \chibar \chi\,,\\
 \Phi =& \psibar \psi + \chibar \chi\,,
\end{align}
and $T^a_f \equiv D f^a$. The physical dimension of these composites are $[\Delta] = [\Phi] = l^{-2}$.

Varying the action with respect to the spinors gives
\begin{align}
 0 =& \epsilon_{abc}e^a e^b \gamma^c D \psi - e^a T_a \psi +\frac{1}{2} \epsilon_{abc}\gamma^c (T^a e^b - e^a T^b)\psi - \frac{1}{2\ell}\epsilon_{abc}e^a e^b f^c \psi\,,\label{diracpsi}\\
 0 =& \epsilon_{abc}e^a e^b \gamma^c D \chi - e^a T_a \chi +\frac{1}{2} \epsilon_{abc}\gamma^c (T^a e^b - e^a T^b)\chi + \frac{1}{2\ell}\epsilon_{abc}e^a e^b f^c \chi\,,\label{diracchi}
\end{align}
where a boundary term has been discarded and we have used the identity
\begin{equation}
 \epsilon_{abc} d(e^a e^b)\gamma^c + \frac{1}{2}\omega^{ab}\epsilon_{cde}e^c e^d\left[\Sigma_{ab},\gamma^e\right] \equiv 0 \,.
\end{equation}

\subsection{Integrability conditions}\label{fieldeqsI}

We will show that integrability conditions on the gravitational equations can be used to derive restrictions on the spinor solutions and in subsection \ref{spinoreqs} we will deal with the spinor equations. 

Note that (\ref{eomf}), which is obtained varying the action with respect to $f^a$ is an algebraic equation for $f^a$, which means that $f^a$is an auxiliary field that contributes no dynamical fields to the system: \eqref{eomf} could in principle be solved algebraically to eliminate $f^a$ from the action. Equation \eqref{eomf} also implies that the classical geometry is conformally related to flat geometries (see below). 

Let us notice that the ansatz $f^a \sim e^a$ and $\Delta = 0$ corresponds to pure MacDowell-Mansouri gravity in $d=4$ \cite{MacDowell:1977jt} and Ach\'ucarro-Townsend gravity in $d=3$ \cite{Achucarro:1986uwr}. We can generalize the ansatz to 
\begin{equation}
 f^a=h(x)e^a\,,\label{genansatz}
\end{equation}
and look for the configurations that comply with Bianchi-identities,
\begin{align}
 D R^{ab} =& 0\,,\label{BianchiDR}\\
 D T^a_f =& R^a{}_b f^b\,.\label{BianchiDTf}
\end{align}
This strategy allows us to restrict the space of solutions by implying relations among $h$, $\Delta$ and $\Phi$ coming from the on-shell restriction of the Bianchi identities. The ansatz (\ref{genansatz}) implies that \eqref{eomf} becomes 
\begin{align}
 R^{ab} = \left(\Phi -\frac{h^2}{\ell^2}\right)e^a\,e^b = \left(\frac{\Phi}{h^2} -\frac{1}{\ell^2}\right)f^a\,f^b  \,, \label{eomf'}
\end{align}
and consequently, both $T^a$ and $T^a_f$ are covariantly constant, $D T^a = R^a{}_b e^b\equiv 0 \equiv R^a{}_b e^b = D T^a_f$. Moreover,
\begin{equation}
 T^a_f= dh\ e^a+h\ T^a \,,\label{Dfrepl}
\end{equation}
so that, for $h \neq 0$ and using eq. (\ref{eomw}), we get
\begin{equation}
%[T^a]_{(\ref{eomw})} = \frac{\ell \Delta}{2h} \epsee - d \ln h e^a\,. \label{Tasol}
T^a = \frac{\ell \Delta}{2h} \epsee - e^a \,d \ln h \,. \label{Tasol}
\end{equation}

Applying exterior covariant derivative to the field equations (\ref{eomf}) and (\ref{eomw}), using the Bianchi identities and replacing (\ref{genansatz}) and (\ref{Tasol}) whenever required, we arrive to the following equations
\begin{align}
 0 &= D\left( \left( \Phi -\frac{h^2}{\ell^2} \right) e^a e^b \right) \,,\\
 0 &= \left( -\frac{1}{2}d\Delta + \frac{\Delta}{h}dh  \right) \epsee \,.
\end{align}

Therefore the conditions
\begin{align}
 \Phi -\frac{h^2}{\ell^2} =& \mathrm{const}\,,\label{condI}\\
 -\frac{1}{2}d\Delta + \frac{\Delta}{h}dh =& 0\label{condII}\,.
\end{align}
are sufficient to satisfy the field equations. Integration of (\ref{condII}) leads to
\begin{equation}
 h^2 = \beta \Delta\,,\label{condIIint}
\end{equation}
where $\beta$ is a constant. Therefore the system is not over constrained so long as the Dirac equation is compatible with conditions (\ref{condI}) and (\ref{condIIint}).

We will show below that an important case is when (\ref{condI}) is set to zero, that implies
\begin{equation}
 \Phi = \frac{\beta}{\ell^2}\Delta\,.\label{condIzero}
\end{equation}

Now, we will study solutions for the spinors in a BTZ background with torsion and $h\ne 0$. Note that the case of constant $\Delta$, $h$ and $\Phi$ corresponds to geometries that admit globally defined Killing spinors \cite{Giribet:2024nwg}, see section \ref{kssection}. The singular case $h=0=\Delta=0$ with $\Phi =$ constant, must be studied separately. This case corresponds to a different sector in which $T^a$ is not determined by \eqref{Dfrepl} and \eqref{Tasol} does not hold. We will come back to this case in the next section\ref{sectionh=0}.

%%%%%%%%%%%%%%%%%%%%%%%%%%%%%%%%%%%%%%
\subsection{The BTZ geometry with torsion as background}    % III . B %
%%%%%%%%%%%%%%%%%%%%%%%%%%%%%%%%%%%%%%

The BTZ geometries are given by the metric
\begin{align}
ds^2= -f^2 dt^2+ \frac{dr^2}{f^2} + r^2(N\,dt + d\varphi)^2 \\
f^2=-M+\frac{r^2}{\ell^2}+\frac{J^2}{4r^2}, \qquad N=-\frac{J}{2r^2}\,,
\end{align}
where the integration constants $M$ and $J$ are the mass and angular momentum and correspond to the conserved charges associated to the global isometries generated by the Killing vectors $\partial_t$ and $\partial_\phi$, respectively. For $M\geq |J|$ these geometries are black holes, while for $M<|J|$ these are naked singularities, except for $M=-1, J=0$, which corresponds to anti-de Sitter spacetime \cite{Banados:1992wn}. The local frames $e^a$ for this metric are 
\begin{align}
e^0&=fdt\,,\label{BTZframes1}\\
e^1&=f^{-1}dr\,,\\
e^2&=r\left( d\varphi +N dt\right)\,,\label{BTZframes-1}\,.
\end{align}
The vanishing of torsion, $de^a+\mathring{\omega }^a{}_b e^{b}\equiv 0$, can be algebraically solved for the connection $\mathring{\omega}^a{}_b$ ,
\begin{align}
\mathring{\omega}^0{}_1 &=\frac{r}{\ell^2}dt-\frac{J}{2r}d\varphi \,,\label{omega-bar1}\\
\mathring{\omega}^1{}_2 &=-f d\varphi \,,\\
\mathring{\omega }^2{}_0&=  -\frac{J}{2f r^2}dr\,.  \label{omega-bar-1}
\end{align}
The corresponding (torsion-free) Riemann curvature two-form is constant, $$\mathring{R}^{ab}=-\ell^{-2} e^a e^b\,.$$

Equation \eqref{eomw} has the form
\begin{equation}
 T_f^a = \tau \epsilon^a{}_{bc} f^b f^c\,, \label{torsionf3d}
\end{equation}
where $\tau = \frac{\ell \Delta}{2h^2}$, which in view of (\ref{condIIint}) must be constant,
\begin{equation}\label{tauconstant}
 \tau = \frac{\ell}{2\beta}\,.
\end{equation}
whose physical units are $[\tau]=l^{-1}$. The specific expression of $\tau$ will not be required in order to solve the spinor equations. 

The most general form for a covariantly constant torsion two-form in three dimensions admits an extra piece,
\begin{equation}
 T_f^a = \tau \epsilon^a{}_{bc} f^b f^c + \lambda(x) f^a\,, \label{torsionpara}
\end{equation}
where $\lambda=\lambda_\mu dx^\mu$ is a scalar one-form. This ``abelian" one-form can be viewed as a Proca field or as an electromagnetic potential, where the freedom to make local rescalings $f^a(x)\to \sigma(x)f^a(x)$ maps into $T^a_f \to \sigma [T^a_f+d \ln \sigma \,f^a]$, $\lambda\to \lambda +d \ln \sigma$ (see e.g., \cite{Andrianopoli:2023dfm}).

From \eqref{torsionf3d}, the contorsion one-form is
\begin{equation}
 \kappa^a{}_c = h \tau  \epsilon^a{}_{bc}e^b\,,\label{kappa}
\end{equation}
and the full Lorentz connection $\omega^a{}_b = \mathring{\omega}^a{}_b+\kappa^a{}_b$ reads
\begin{align}
\omega^{01}& =\frac{1}{r f}\left(\frac{r^2}{\ell^2}-\frac{J^2}{4 r^2}\right)e^0+\frac{1}{2} \left(2h \tau  -\frac{J}{r^2}\right)e^2 \,,\label{omega01}\\
\omega^{02} & = -\frac{1}{2}\left(2h \tau + \frac{J}{r^2}\right)e^1\,,\label{omega02}\\
\omega^{12} & = -\frac{1}{2}\left(2 h \tau + \frac{J}{r^2}\right)e^0-\frac{f}{r} e^2\,,\label{omega12}
\end{align}
With this connection one can explicitly check that the curvature takes the form \eqref{eomf}, \begin{equation}
R^{ab} = - \left( \frac{1}{\ell^2} - h^2 \tau^2 \right)e^a e^b\,.
\end{equation}
Other three-dimensional black hole solutions in the presence of torsion have also been found in \cite{Mielke:1991nn,Blagojevic:2003vn,Garcia:2003nm}.

%%%%%%%%%%%%%%%%%%%%%%%%%%%%%%%%%
\section{Solving the spinor equations}\label{spinoreqs}
%%%%%%%%%%%%%%%%%%%%%%%%%%%%%%%%%

Following the ideas in \cite{Alvarez:2022bqe}, our strategy is to impose the integrability constraints in the previous section to look for solutions to the spinor equation.

Before proceeding further let us recall the spinor conventions that we adopted in this paper. The Dirac adjoint was defined in the Introduction section and $(\gamma^a)^\dagger = \gamma^0 \gamma^a \gamma^0$. The $\gamma$-matrices satisfy $\{\gamma^a,\gamma^b\}=2 \eta^{ab}$, where the metric $\eta$ is given by $\eta=\mathrm{diag}(-,+,+)$, and the spinor indices are often omitted. Therefore the spinor bilinear have the following reality conditions
\begin{align}
 (\psibar \psi)^\dagger =& \psibar \psi\,,\\
 (\psibar \gamma^a\psi)^\dagger =& -\psibar \gamma^a\psi\,.
\end{align}

When looking for solutions it is also useful to adopt the spinor field equations, \eqref{diracpsi} and \eqref{diracchi}, in terms of tensorial notation,
\begin{align}
 0 =& \gamma^\mu D_\mu \psi + \frac{3}{4} \epsilon_a{}^{bc} (De^a)_{bc}\psi - \frac{3\theta h}{2\ell} \psi\,,\\
 0 =& \gamma^\mu D_\mu \chi + \frac{3}{4} \epsilon_a{}^{bc} (De^a)_{bc}\chi + \frac{3\theta h}{2\ell} \chi\,,
\end{align}
where $De^a = (1/2) (De^a)_{bc} e^b e^c$ and $\theta=\pm 1$ corresponds to the two possible choices of inequivalent representations of the Clifford algebra in odd dimensions.

%%%%%%%%%%%%%%%%%%%%%%%%%%%%%%%%%%%%%%%%%%%%%%%%%%%%%%%%%
\subsection{Dirac equation in reduced generic background}
%%%%%%%%%%%%%%%%%%%%%%%%%%%%%%%%%%%%%%%%%%%%%%%%%%%%%%%%%

After separating the torsionless and torsion-full components, and using \eqref{torsionf3d}, the Dirac equations of the model are given by
\begin{align}
 0 = \gamma^\mu \mathring{D}_\mu \psi -\left[\frac{3\theta h}{2\ell}  + 9\left(1-\frac{\theta}{6}\right)h \tau \right]\psi\,,\label{diracpsitensor}\\
 0 = \gamma^\mu \mathring{D}_\mu \chi +\left[\frac{3\theta h}{2\ell}  - 9\left(1-\frac{\theta}{6}\right)h \tau \right]\chi\,,\label{diracchitensor}
\end{align}
where
\begin{equation}
 \mathring{D}_\mu \psi = \partial _\mu \psi +\frac{1}{2} \mathring{\omega}^{ab}{}_\mu\Sigma_{ab}\psi-\frac{i}{2} A^I{}_\mu \sigma_I \psi \,.
\end{equation}
Here $\tau$ should be treated as an arbitrary  function of the coordinates, but consistency requires (\ref{condI}) and (\ref{condIIint}) to be checked once the solution for $\psi$ and $\chi$ are obtained.

The explicit form of the equations, when $A^I = 0$ and $B^I = 0$, is
\begin{align}
 0 &= \gamma^0\left[\frac{1}{f}\partial_t + \frac{J}{2r^2 f}\partial_\varphi\right] \psi + \gamma^1(f \partial_r + A )\psi + \gamma^2 \frac{1}{r}\partial_\varphi \psi + B \psi\,,\\
 0 &= \gamma^0\left[\frac{1}{f}\partial_t + \frac{J}{2r^2 f}\partial_\varphi\right]\chi +   \gamma^1(f \partial_r + A)\chi + \gamma^2 \frac{1}{r}\partial_\varphi \chi  + C \chi\,,
\end{align}
(here $\gamma^0=\gamma^a|_{a=0}$, etc...)
Assuming only radial dependence $\psi=\psi(r)$ and $\chi=\chi(r)$, we get 
\begin{align} \label{psi-r}
 0 &= \gamma^1(f \partial_r + A )\psi + B \psi \,,\\
 \label{chi-r}
 0 &= \gamma^1(f \partial_r + A)\chi + C \chi \,,
\end{align}
where
\begin{align}
 A =& -\frac{M - 2(r/l)^2}{2r f}\,,\\
 B =& \frac{\theta J}{4r^2} -\frac{3\theta h}{2\ell}  - 9\left(1-\frac{\theta}{6}\right)h \tau \,,\\
 C =& \frac{\theta J}{4r^2} +\frac{3\theta h}{2\ell}  - 9\left(1-\frac{\theta}{6}\right)h \tau\,.
\end{align}
Using the projectors
\begin{equation}
 \gamma_\pm = \frac{1}{2}(1 \pm \gamma^1)\,,
\end{equation}
we can rewrite (\ref{psi-r}, \ref{chi-r}) as
\begin{align}
 0 &= \gamma_+ (f \partial_r + A + B)\psi + \gamma_- (-f \partial_r - A + B)\psi\,,\\
 0 &= \gamma_+ (f \partial_r + A + C)\chi + \gamma_- (-f \partial_r - A + C)\chi\,. 
\end{align}

Splitting $\psi$ and $\chi$ into their $\pm$ projections, $\psi = \psi_+ + \psi_-\,, \quad \chi = \chi_+ + \chi_-$, with $\gamma_\pm \psi_\pm = \psi_\pm$, $\gamma_\pm \psi_\mp = 0$, the Dirac equation is easily integrated as 
\begin{align}
 \psi_+ & = \rho_+ e^{i\alpha_+}\exp \left( -\int dr \ f^{-1}  (A + B) \right)  u_+\,,\label{sol1}\\
 \psi_- & = \rho_- e^{i\alpha_-}\exp \left( -\int dr \ f^{-1}  (A - B) \right)  u_-\,,\\
 \chi_+ & = \sigma_+ e^{i\beta_+}\exp \left( -\int dr \ f^{-1}  (A + C) \right)  u_+\,,\\
 \chi_- & = \sigma_- e^{i\beta_-}\exp \left( -\int dr \ f^{-1}  (A - C) \right)  u_-\,,\label{sol-1}
\end{align}
where $\rho_\pm$, $\sigma_\pm$, $\alpha_\pm$ and $\beta_\pm$ are real constants and $u_\pm$ are normalized constant spinors,
\begin{equation}
 u_\pm = \left(\begin{array}{c}
          1/\sqrt{2}\\
          \pm 1/\sqrt{2}
         \end{array}\right)
\end{equation}

Since the gravitating solution requires $\Phi \propto \Delta$, the following properties are crucial: the projectors $\gamma_\pm$ are Hermitian and $\gamma_\pm \gamma^0 = \gamma^0 \gamma_\mp$. Indeed, thanks to these properties, the $B$ and $C$ functions in (\ref{sol1})-(\ref{sol-1}) cancel out from (\ref{psibarpsi}) and (\ref{chibarchi}). Therefore, when computing the bilinears we obtain
\begin{align}
 \psibar \psi &= 2\rho_+ \rho_- \sin(\alpha_+ - \alpha_-)\exp \left( -2 \int dr \ f^{-1}  A \right)\,, \label{psibarpsi}\\
 \chibar \chi &= 2\sigma_+ \sigma_- \sin(\beta_+ - \beta_-) \exp \left( -2 \int dr \ f^{-1}  A \right)\,.\label{chibarchi}
\end{align}
and therefore $\Phi$ and $\Delta$ are proportional without imposing further restrictions on $h$. Upon integration, the exponential term is found to be $1/(2\ell r f)$ that can be written expressed as
\begin{equation}
 \exp \left( -2 \int dr \ f^{-1}  A \right) = \frac{1}{2\sqrt{r_+^2 - r_-^2}}\left( \frac{1}{r^2 - r_+^2} - \frac{1}{r^2 - r_-^2} \right)^{1/2}\,,
\end{equation}
which blows up at $r_{\pm}$. Therefore 
\begin{equation}
 \Delta = \frac{\delta}{2lr f}\,,\label{Deltadelta}
\end{equation}
where $\delta$ is a constant that depends solely on the integration constants of the spinor solution
\begin{equation}
 \delta = 2(\rho_+ \rho_- \sin(\alpha_+ - \alpha_-) - \sigma_+ \sigma_- \sin(\beta_+ - \beta_-))\,.
\end{equation}

%%%%%%%%%%%%%%%%%%%%%%%%%%%%%%%%%%%%%%%%%%%%%%%%%%%
\subsubsection{Energy condition}\label{energycond}
%%%%%%%%%%%%%%%%%%%%%%%%%%%%%%%%%%%%%%%%%%%%%%%%%%%

In three dimensions we define the energy momentum 2-form by
\begin{equation}
 \delta L = \delta e^a \tau_a\,,
\end{equation}
where
\begin{equation}\label{taud}
 \tau_a=\frac{1}{2}\tau_a{}^\mu\epsilon_{\mu bc} e^b e^c\,.
\end{equation}

The variation of (\ref{spinorlag}) with respect to $e^a$ gives us
\begin{align}
 \delta_e L=& \delta e^a \kappa \left[
 \epsilon_{abc}\psibar(e^b\gamma^c D+\overleftarrow{D}e^b\gamma^c)\psi
 -\epsilon_{abc}\chibar(e^b\gamma^c D+\overleftarrow{D}e^b\gamma^c)\chi \nonumber \right.\\
&\left.-2T_a\Delta-\frac{1}{\ell}\epsilon_{abc}e^bf^c \Phi-i\epsilon_{abc}e^b B_I(\psibar\gamma^c\sigma^I\psi + \chibar\gamma^c\sigma^I\chi)+e_a d\Delta\right]\,,
\end{align}
plus a boundary term (given by $-\kappa d(\delta e^a e_a \Delta)$). The terms proportional to $T^a$ and $d\Delta$ combine in the following way
\begin{equation}
\delta e^a \tau_a (T^a,d\Delta) = \delta e^a \kappa (-2T_a \Delta + e_a d\Delta)\,,
\end{equation}
using $T^a$ from (\ref{Tasol}), (\ref{condII}) and the definmition of $\tau_{ab}$ implied from (\ref{taud}), we get
\begin{equation}
\tau_{ab}(T^a,d\Delta) = -4\kappa \ h \tau \Delta \eta_{ab} \,.
\end{equation}

Using also (\ref{tauconstant}) we get,
\begin{equation}\label{tauabTa}
 \tau_{ab} (T^a, d\Delta)= -2\kappa\frac{ l \Delta^2}{h} \eta_{ab} \,,
\end{equation}

Therefore
\begin{align}
 \tau_{ab} =& \kappa \left[ \eta_{ab} \psibar (\gamma^c D_c -\lD_c\gamma^c ) \psi - \half \psibar\left( \gamma_a D_b +\gamma_b D_a \right)\psi + \half \psibar\left( \lD_b\gamma_a  +\lD_a\gamma_b \right)\psi\right.\nonumber\\
 &-\eta_{ab} \chibar (\gamma^c D_c -\lD_c\gamma^c ) \chi + \half \chibar\left( \gamma_a D_b +\gamma_b D_a \right)\chi - \half \chibar\left( \lD_b\gamma_a  +\lD_a\gamma_b \right)\chi \nonumber\\
 &\left.-2\left(\frac{ \ell \Delta^2}{h} + \frac{h}{l}\Phi\right)\eta_{ab}\right]\,.\label{tau_ab}
\end{align}

Finally, replacing solutions (\ref{sol1})-(\ref{sol-1}) in (\ref{tau_ab}) leads us to
\begin{align}
 \tau_{ab} =& \kappa \left(
\begin{array}{lll}
-\frac{1}{2}x + \frac{\Delta J}{2r^2} & 0 & \frac{\Delta(J^2-2Mr^2)}{4fr^3}\\
0 & -x & 0\\
\frac{\Delta(J^2-2Mr^2)}{4fr^3} & 0 & \frac{1}{2}x + \frac{\Delta J}{2r^2}
                     \end{array}\right)
\,.\label{tau_abmatrix}
\end{align}
where
\begin{equation}
 x = \frac{\ell}{h}\left( \Delta^2 + 2 \frac{h^2 \Phi}{\ell^2}\right)\,.
\end{equation}

We can see that the tensor (\ref{tau_abmatrix}) is symmetric and traceless,
\begin{equation}
 \tau_a {}^a = 0\,.\label{traceless}
\end{equation}
The reason behind this property is the intrinsic Weyl-rescaling invariance of the action
\begin{align}
 e^a &\rightarrow \lambda e^a\\
 \psi &\rightarrow \lambda^{-1} \psi\\
 \psibar &\rightarrow \lambda^{-1} \psibar\\
 \chi &\rightarrow \lambda^{-1} \chi\\
 \chibar &\rightarrow \lambda^{-1} \chibar\,,
\end{align}
where $\lambda = \lambda(x)$, together with the fact that the fermionic piece of the action has well defined scaling weights $(2,1,1,1,1)$ with respect to the fields $(e^a,\psi, \psibar, \chi, \chibar)$. The latter implies
\begin{equation}
 e^a \frac{\delta L_\mathrm{fer}}{\delta e^a} + \psibar \frac{\delta L_\mathrm{fer}}{\delta \psibar} + \frac{\delta L_\mathrm{fer}}{\delta \psi}\psi + \chibar \frac{\delta L_\mathrm{fer}}{\delta \chibar} + \frac{\delta L_\mathrm{fer}}{\delta \chi}\chi = 4L_\mathrm{fer}\,,\label{Lhom}
\end{equation}
where $L_\mathrm{fer} = L_\psi^+-L_\chi^-$. But it is also valid that the corresponding pieces of the action are proportional to the spinor field equations
\begin{align}
 \psibar \frac{\delta L_\psi^+}{\delta \psibar} + \frac{\delta L_\psi^+}{\delta \psi}\psi =& L_\psi^+\\
 \chibar \frac{\delta L_\chi^-}{\delta \chibar} + \frac{\delta L_\chi^-}{\delta \chi}\chi =& L_\chi^-
\end{align}
and therefore the fermionic action vanishes on-shell, $L_\mathrm{fer} \approx 0$. Using this in (\ref{Lhom}) we demonstrate that the traceless condition, (\ref{traceless}), must be satisfied for any configuration that ought to be a solution of the fermionic equations.

From (\ref{Deltadelta}) and the integrability conditions (\ref{condIIint}) and (\ref{condIzero}) we obtain the asymptotic behavior of $\Delta$, $\Phi$ and $h$
\begin{align}
 \Delta &\sim \frac{\delta}{2\ell ^2} \left( \rlfrac^{-2} + \frac{M}{2} \rlfrac^{-4} - \frac{J^2}{8r^2}\rlfrac^{-4}\right) =\frac{\delta}{2} r^{-2} + O (r^{-4})\,,\label{falloff1}\\
 \Phi &\sim \frac{\beta}{\ell^2} \Delta = \frac{\beta\delta}{2\ell^2} r^{-2} + O (r^{-4})\,,\\
 h^2 &\sim \beta \Delta = \frac{\beta\delta}{2} r^{-2} + O (r^{-4})\,.\label{falloff3}
\end{align}
From (\ref{falloff1})-(\ref{falloff3}), we can determine the fall-off of the functions in the stress energy tensor

\begin{align}
 x & \sim \left( \frac{\delta^3}{8\ell^4\beta} \right)^{1/2}\left( 1+\frac{2\beta^2}{\ell^4} \right)
 \rlfrac^{-3}\left(1  + \left(\frac{3M}{4}  - \frac{3J^2}{16r^2}\right)\rlfrac^{-1}\right) \,,\\
 \frac{\Delta J}{2r^2} & \sim \left( \frac{\delta J}{4\ell^4} \right)
 \rlfrac^{-4}\left( 1 + \left(\frac{M}{2}  - \frac{J^2}{8r^2}\right)\rlfrac^{-2}\right) \,,\\
 \frac{\Delta(J^2-2Mr^2)}{4fr^3} &\sim -\frac{\delta}{4\ell^3}
 \rlfrac^{-4}\left( 1 + \left(M - \frac{J^2}{4r^2}\right)\rlfrac^{-2}\right)\left(M- \frac{\delta J^2}{2\ell^2}\rlfrac^{-2}  \right)\,,
\end{align}
therefore
\begin{align}
 \tau_{00} \sim \tau_{00}^{(0)} r^{-3}+O(r^{-4})\,,\\
 \tau_{11} \sim \tau_{11}^{(0)} r^{-3}+O(r^{-4})\,,\\
 \tau_{22} \sim \tau_{22}^{(0)} r^{-3}+O(r^{-4})\,,\\
 \tau_{01} \sim \tau_{01}^{(0)} r^{-4}+O(r^{-6})\,.
\end{align}
These fall-off conditions guarantee finiteness in the contribution to the charges at spatial infinity \cite{Brown:1986nw}.

In AdS$_3$ Chern-Simons gravity is convenient to compute the charges using the term provided in \cite{Miskovic:2006tm},
\begin{equation}
 Q(\xi) = \frac{k}{4\pi} \int_{\partial \Sigma} Tr(A i_\xi A)\,,\label{MiskovicOlea}
\end{equation}
where $k/(4\pi)$ is the global constant in front of the Chern-Simons action, $\xi$ is a Killing vector and $\partial \Sigma$ is a 1-dimensional boundary circle at constant $t$ and $r \rightarrow \infty$. When using (\ref{MiskovicOlea}) for the action (\ref{action}), we have $k/(4\pi) = 2\ell$ and we will get contributions from the bosonic and fermionic parts of $\AAA_+$ and $\AAA_-$.

For a time-like Killing vector one gets
\begin{equation}
 Q(\partial_t) = 2\ell\int_{S^1} d\phi \ ( (\omega_+)^{ab}{}_\phi (\omega_+){}_{ab}{}_t - (\omega_-)^{ab}{}_\phi (\omega_-)_{ab}{}_t) -4\ell\int_{S^1} d\phi \ e^a{}_\phi e_{at} \Delta\,,
\end{equation}
that leads to
\begin{equation}
 Q(\partial_t) = \lim_{r\rightarrow \infty}\left(-16\pi h (M - h\tau J) + 4\pi\ell J \Delta \right)\,,
\end{equation}
which vanishes because of the asymptotics (\ref{falloff1})-(\ref{falloff3}).

For a space-like Killing vector one gets
\begin{equation}
 Q(\partial_\phi) = 2\ell\int_{S^1} d\phi \ ( (\omega_+)^{ab}{}_\phi (\omega_+)_{ab}{}_\phi - (\omega_-)^{ab}{}_\phi (\omega_-)_{ab}{}_\phi) -4\ell\int_{S^1} d\phi \ e^a{}_\phi e_{a\phi} \Delta\,,
\end{equation}
that leads to
\begin{equation}
 Q(\partial_\phi) = \lim_{r\rightarrow \infty}\left(16\pi (h J - ( 2\beta \tau + \ell) r^2 \Delta \right)\,,
\end{equation}
and therefore a non-vanishing finite charge:
\begin{equation}
 Q(\partial_\phi) = -16\pi\ell \delta \,.
\end{equation}

The purely bosonic BTZ configuration with torsion can not be continuously recovered by setting the fermionic fields to zero, however, the torsionless BTZ background is recovered by in the singular limit
\begin{equation}
\left.
\begin{array}{lrr}
        \Delta &\rightarrow & 0\,\\
\Phi &\rightarrow & 0\,\\
\beta &\rightarrow & \infty
\end{array}\right\} \quad \text{keeping }  h \text{ const}\,.
\end{equation}

The fact that the bosonic BTZ can be reached by an improper limit only indicates that the present solution does not belong to the class of fermionic wigs discussed by Gentile et al \cite{Gentile:2012tu}.

%%%%%%%%%%%%%%%%%%%%%%%%%%%%%%%%%%%%%%%%%%%%%
\subsection{Covariantly constant spinors: BPS geometries}\label{kssection}  % 4 %
%%%%%%%%%%%%%%%%%%%%%%%%%%%%%%%%%%%%%%%%%%%%%

We will now discuss some cases in which the field equations for the fermionic field can be fully integrated. This occurs for certain particular values of $M$ and $J$ that correspond to geometries that admit globally defined covariantly constant (Killing) spinors.

We first note that since subalgebras $su(2|2)_+$ and $su(2|2)_-$ commute, the levels $\kappa_{\pm}$ in each of the Lagragians \eqref{action} need not be the same and one could have equally taken any linear combination $\alpha L_\text{CS}(\AAA_+)- \beta L_\text{CS}(\AAA_-)$. Additionally, if we are only interested in the system containing one copy of $SU(2)\times U(1)$, the relevant Lagrangian would be
\begin{equation}
L_{Eff} = L_\text{SO(2,2)}(\omega, f)+L_{\text{SU(2)}}(\AAA)+L_{\text{F}}(\psi) \,,
\end{equation}
where 
\begin{align}
L_\text{SO(2,2)}(\omega, f) &= \frac{1}{G} \epsilon_{abc} (R^{ab} +\frac{1}{3\ell^2} f^a\, f^b)\, f^c   \\
L_{\text{SU(2)}}(\AAA) &= \kappa Tr[\AAA d\AAA + \frac{2}{3}\AAA^3]\\
L_{\text{F}}(\psi) &=\frac{\kappa}{2}(\overline{\psi}\se )(\overleftarrow{\nabla }-\nabla )(\se\psi) + i\frac{\kappa}{2}\overline{\psi}\se A^I\sigma_I \se\psi\\
& =\frac{\kappa}{2}\overline{\psi}(\se\se \nabla -\overleftarrow{\nabla }\se\se )\psi-\frac{\kappa}{2}\overline{\psi}(\se\slashed{T}+\slashed{T}\se -\frac{1}{\ell}\se\slashed{f}\se+ i\se A^I\sigma_I \se )\psi\,\,,
\end{align}
where $\nabla $ is the covariant derivative for the AdS connection acting on the fermion,
\begin{equation}
\nabla \psi := (d+\frac{1}{2} \omega^{ab} \JJ_{ab} + \frac{f^a}{\ell}  \PP_a )\psi \,.
\end{equation}
Here we take $\JJ_{ab}=\frac{1}{2}\gamma_{ab}=\Sigma_{ab}$, $\PP_a= \gamma_a$, and
\begin{equation}
 \gamma^0 = i \theta \sigma_1\,, \quad \gamma^1 = \theta \sigma_3\,, \quad \gamma^2 = \theta \sigma_2\,.
\end{equation}

As shown in \cite{Giribet:2024nwg}, there exist a family of 2+1 locally AdS geometries that admit covariantly constant spinor fields, that is, spinors $\xi$ that solve
\begin{equation} \label{K-spinor}
\nabla_\text{(adS)} \ \epsilon = d\epsilon +\frac{1}{2} \omega^{ab}\Sigma_{ab} \epsilon +\frac{1}{\ell }\,e^a \gamma_a  \epsilon =0\,.
\end{equation}
Here
\begin{equation}
 \gamma^0 = -\theta i \sigma_2\,, \quad \gamma^1 = \theta \sigma_1\,, \quad \gamma^2 = \theta \sigma_3\,, \quad 
\end{equation}
Equation \eqref{K-spinor} separates is easily handled if one separates left and right mover modes by introducing coordinates $x^{\pm} =t\pm \theta \ell \phi$, $\partial _{\pm } = \frac{1}{2\ell }\,\left(\ell \partial _{t}\pm \theta \partial_{\phi }\right)$. Note that a left-mover with a given value of $\theta$ is equivalent to a right-mover for $-\theta$.

The component of \eqref{K-spinor} proportional to the one-form $dx^-$ is simply $\partial_- \epsilon= 0$ and therefore $\epsilon=\epsilon(x^+,r)$. Following \cite{Giribet:2024nwg}, the solutions of \eqref{K-spinor} for generic $(J, M)$ take the form
\begin{equation}
\epsilon = U \left(\eta_1 e^{i\frac{\omega}{2\ell }\,x^+ } u_+ + \eta_2 e^{-i\frac{\omega}{2\ell }\,x^+ } u_- \right)\,, \label{ks}
\end{equation}
where
\begin{equation}
 U = X^{-1} \gamma_+ + X \gamma_-\,, \quad X = \left( \frac{r}{\ell} -\frac{\theta J}{2r} + f\right)^{1/2}\,,
\end{equation}
\begin{equation}
 u_\pm =  \left(\begin{array}{c}
1 \\
-\theta \frac{1 \pm i\omega}{1 \mp i\omega}
\end{array}\right) \,, \quad \omega = \sqrt{\frac{\theta J}{\ell} - M}
\end{equation}
and $\eta_1$, $\eta_2$ are arbitrary normalization constants for the right- and left-moving spinors. $u_\pm$ are two component spinors with fixed components that play the role of normal modes of the equation for the $dx^-$ component in \eqref{K-spinor}. The form of the operator $U$ and $X=X(r)$ is determined by the radial component equation coming from \eqref{K-spinor}. The constant $\omega$ involves parameters that define the background geometry and the sign of the $\gamma$-matrices representation.

In order for the spinors \eqref{ks} to be globally well defined,  they should be either periodic or antiperiodic for $\phi \rightarrow \phi + 2\pi$, which leads to the condition
\begin{equation}
\omega = n \in \mathbb{Z}_{\geq 0}\quad \Rightarrow \quad \frac{\theta J}{\ell} - M = n^2 \ge 0\,. \label{global}
\end{equation}

An important feature of solutions \eqref{ks} is that they have a constant finite norm\footnote{In order to see how this works let us use the fact that
\begin{equation}
 U^\dagger \gamma^0 U = 2\gamma^0\,,
\end{equation}
and
\begin{equation}
 u^\dagger_\pm \gamma^0 u_\pm = \pm \frac{4i \omega}{1 + \omega^2}\,, \qquad u^\dagger_\pm \gamma^0 u_\mp = 0\,,
\end{equation}
}
,
\begin{equation}
 \bar{\epsilon}\epsilon= -\frac{8\omega}{1 + \omega^2}(|\eta_1|^2 + |\eta_2|^2)\,,
\end{equation}
and does not depend on the value of $\theta$ either.\footnote{The extremal black hole case $J=\pm M$ ($\alpha=0$) leads to a spinor of vanishing norm.}  This fact can be exploited to construct solutions to the Dirac equations \eqref{diracpsitensor} and \eqref{diracchitensor}. Consider two spinors $\psi$ and $\chi$ of the form \eqref{ks},
\begin{align}
 \psi =  U \left(\eta^{(\psi)}_1 e^{i\frac{\omega}{2\ell }\,x^+ } u_+ + \eta^{(\psi)}_2 e^{-i\frac{\omega}{2\ell }\,x^+ } u_- \right)\,, \label{ks-psi}\\
 \chi =  U \left(\eta^{(\chi)}_1 e^{i\frac{\omega}{2\ell }\,x^+ } u_+ + \eta^{(\chi)}_2 e^{-i\frac{\omega}{2\ell }\,x^+ } u_- \right)\,, \label{ks-chi}
\end{align}
whose respective norms are
\begin{align}
 \Delta =  \frac{8\omega}{1 + \omega^2}(|\eta^{(\psi)}_1|^2 + |\eta^{(\psi)}_2|^2 - |\eta^{(\chi)}_1|^2 - |\eta^{(\chi)}_2|^2)\,,\label{Deltaks}\\
 \Phi =  \frac{8\omega}{1 + \omega^2}(|\eta^{(\psi)}_1|^2 + |\eta^{(\psi)}_2|^2 + |\eta^{(\chi)}_1|^2 + |\eta^{(\chi)}_2|^2)\,.
\end{align}

Since $\Delta$ and $\Phi$ are constants, so are $h$ and $\tau$ as well. In other words, the back reaction of these covariantly constant spinors does not appreciably the geometry, which remains a constant negative curvature and constant torsion manifold.  An important point to note is that in order for $\psi$ and $\chi$ to be solutions of the corresponding Dirac equations,
\begin{align}
 0 =& \theta\epsilon_{abc}e^a e^b \gamma^c \mathring{D} \psi +\frac{1}{2} e^a T_a \psi +\frac{1}{2}\psibar \epsilon_{abc}\gamma^c (T^a e^b - e^a T^b)\psi - \frac{\theta}{2\ell}\epsilon_{abc}e^a e^b f^c \psi\,,\label{diracpsitheta}\\
 0 =& \theta\epsilon_{abc}e^a e^b \gamma^c D \chi +\frac{1}{2} e^a T_a \chi +\frac{1}{2}\chibar \epsilon_{abc}\gamma^c (T^a e^b - e^a T^b)\chi + \frac{\theta}{2\ell}\epsilon_{abc}e^a e^b f^c \psi\,.\label{diracchitheta}
\end{align}
additional conditions are imposed on the normalization constants. This is due to the fact that $\Delta$ and $\Phi$ enter in \eqref{diracpsitheta} and \eqref{diracchitheta} through the torsion.

Assuming $\psi$ and $\chi$ to be two independent solutions, requires
\begin{align}
 0 = \frac{\theta}{\ell} + \frac{h \tau}{2} + \frac{\theta h}{2\ell} \,,\label{diracpsithetacond}\\
 0 = \frac{\theta}{\ell} + \frac{h \tau}{2} - \frac{\theta h}{2\ell} \,,\label{diracchithetacond}
\end{align}
where $h$ and $\tau$ depend on $\Delta$ and $\Phi$ via \eqref{tauconstant} and \eqref{condIIint}. It is clear that (\ref{diracpsithetacond},\ref{diracchithetacond}) are too strong, implying $h=0$ and $1/\ell =0$. Alternatively, taking $\chi=0$ ($\Delta=\Phi$), or $\psi=0$ ($\Delta=-\Phi$). The two cases $\Delta=\pm \Phi$ correspond to $\beta=\pm\ell^2$ and therefore $h^2=\pm \ell^2 \Delta$ and $\tau=\pm(2\ell)^{-1}$, which reduces to the condition
\begin{equation}
 \Delta = \frac{\mp 16/\ell^2}{1+2\theta^2}\,.\label{Deltaspinorsolution}
\end{equation}
The value of the bilinear in \eqref{Deltaspinorsolution} cannot be continuously deformed to the purely bosonic solution which means that this class of solutions does not correspond to a fermionic wig \cite{Gentile:2012tu}. In the context of black holes, a ``hair'' is an attribute of a black hole in addition to its mass, angular momentum and electric charge. This feature of a black hole solution should be characterized by a parameter that can be continuously set to zero to connect with a generic black hole without hair. In this sense, the spinor field configuration \eqref{ks-psi},  \eqref{ks-chi} (subjected to \eqref{Deltaspinorsolution}) does not qualify as black hole hair (or ``wig").

One could expect that, since the constant spinor condition $\nabla \psi=0$ is stronger than the Dirac equation $\gamma^\mu \nabla_\mu \psi=0$, there could be more solutions satisfying the latter. It is important to clarify that the covariantly constant solution exists for very special values of $M,J$, not for a generic spherically symmetric stationary geometry. In addition, the solution includes the full non perturbative back reaction on the geometry, its amplitude is fixed and cannot be continuously switched off. Another difference between the solutions in sections 4.1 and 4.2 lies in the fact that the constant spinors exist only for BTZ geometries with $M<|J|$, which correspond to naked singularities and not black holes. These naked singularities can be of the overspinning type ($|J|>|M|$) \cite{Briceno:2021dpi} or spinning angular excesses $M \leq -|J|$ \cite{Giribet:2024nwg}.

Finally, the case $h=0$ must be treated separately, as mentioned in subsection \ref{fieldeqsI}. In this case the Bianchi identity \eqref{BianchiDTf} implies $\tau = 0$, and therefore equation \eqref{eomw} implies $\Delta = 0$. In the search of nontrivial spinor solutions, we can now focus on the case when \eqref{condI} is a constant different from zero. This implies that $\Phi = \text{const} \ne 0$. Conditions \eqref{diracpsithetacond} and \eqref{diracchithetacond} are still valid replacing $h\rightarrow 0 $ and therefore these conditions lead to $\theta/\ell =0$, implying that the case $h=0$ is only compatible with the trivial solution.

%%%%%%%%%%%%%%%%%%%%%%%%%%%%%%%%%%%%%%%%%%%%%%%%%%%%%%%
\section{Discussion and conclusions}\label{sec:conclu}
%%%%%%%%%%%%%%%%%%%%%%%%%%%%%%%%%%%%%%%%%%%%%%%%%%%%%%%

In this work, we have presented exact analytical solutions describing gravitating spinor fields on a BTZ black hole background with torsion, within the framework of a three-dimensional Chern–Simons gauge theory for unconventional supersymmetry. This model, characterized by matter in the adjoint representation and no requirement for fermion–boson degree-of-freedom matching, permits a consistent inclusion of spinor fields as components of the gauge connection.

Our analysis shows that nontrivial spinor configurations can consistently exist on a torsional BTZ geometry. These spinor solutions satisfy a set of integrability conditions derived from the Bianchi identities, and the resulting geometry exhibits a non-vanishing torsion sourced by bilinear spinor terms. Despite the divergence of the spinor field near the black hole horizon, and the corresponding energy–momentum tensor divergence the conserved charges remain finite.

We have also used the existence of covariantly constant spinors in adS background to construct spinor solutions. The former require special conditions on the background parameters (such as a quantization condition on mass and angular momentum), that have to be supplemented with extra conditions on the integration constants of the spinors in order to construct new classes of solutions to the spinor equations. These classes demand non-zero values of the spinor bilinears, and therefore they are not connected continuously to the purely bosonic sector. The distinction from a fermionic wig highlights the novel character of these configurations.

Looking ahead, several directions are worth exploring. One natural extension is the investigation of supersymmetric solutions in higher-dimensional USUSY models and the role of torsion in the presence of matter couplings beyond the spin-1/2 sector. Furthermore, studying perturbations around the spinor–torsion background could shed light on its stability and possible implications for holography in $AdS_3/CFT_2$. Finally, understanding the quantization and physical interpretation of the conserved charges associated with the spinor sector may provide further insights into the structure of supersymmetric gravity theories in lower dimensions.

\section{Acknowledgements}

This work has been funded in part with support from ANID-FONDECYT grants 1220862, 1230112, and 1241835. We thank G. Giribet, C. Mart\'inez, O. Mi\v{s}kovi\'c, for valuable discussions and insightful comments during various stages of this work. J.O. acknowledges the hospitality of the Centro de Estudios Científicos (CECs), where part of this work was carried out.

\begin{appendices}

\section{$su(2|2)$ representation}\label{apprep}

We used the following representation of $su(2|2)$ for the ``$+$'' and ``$-$'' sectors of the algebra:
\begin{eqnarray}
&\JJ_a =\left[\begin{array}{c|c}
\frac{1}{2}\gamma_a &  0_{2\times2}\\[0.5em] \hline
0_{2\times2} & 0_{2\times2} \\
\end{array}\right]\,, \quad \text{or} \quad (\JJ_a)^A_{\ B}=\frac{1}{2}(\gamma_a)^A_{\ B}\,,&\\
&\TT_{I} =\left[\begin{array}{c|c}
0_{2\times2} &  0_{2\times2}\\[0.5em] \hline
0_{2\times2} & -\frac{i}{2} (\sigma_I)^t \\
\end{array}\right]\,, \quad \text{or} \quad (\TT_I)^A_{\ B}=-\frac{i}{2}(\sigma_I^t)^A{}_B\,,&\\
&(\QQ^\alpha_i)^A_{\ B}=\left[\begin{array}{c|c}
0_{2\times2} & 0_{2\times 2}\\ [0.5em] \hline
\delta^A_i \delta^\alpha_B & 0_{2\times2}
\end{array}\right]\,,&\\
&(\overline{\QQ}_\alpha^i)^A_{\ B}=\left[\begin{array}{c|c}
0_{2\times2} & \delta^A_\alpha \delta^i_B\\ [0.5em] \hline
0_{2\times 2} & 0_{2\times2}
\end{array}\right]\,,&\\
&\mathbb{Z}^A_{\ B}=\left[\begin{array}{c|c}
\frac{i}{2}\delta^\alpha_\beta &0_{2\times2}\\ [0.5em] \hline
0_{2\times2} & \frac{i}{2}\delta^i_j\end{array}\right]=\frac{i}{2}(\delta^A_\alpha \delta^\alpha_B+\delta^A_i \delta^i_B)\,.&
\end{eqnarray}
The $\gamma$-matrices are in a $2\times 2$ spinorial-representation with indices $\alpha, \beta=1,2$. The indices of the tangent space $a,b=0,1,2$. Indices in the adjoint representation of $SU(2)$ take values $I,J=1,2,3$ and indices in the fundamental take values $i,j=1,2$. We chose the upper-left block of the representation for spinor indexes and lower-right block for the indexes in the fundamental representation of $SU(2)$, so the split $A=(\alpha,i)$.

We have two complex spinors $\psi^\alpha_i$. The Pauli matrices satisfy $[\sigma_I,\sigma_J]=2i\epsilon_{IJ}^{\ \ \ K} \sigma_K$.

Using the fact that in three dimensions the $\gamma$-matrices satisfy $\gamma_a \gamma_b \gamma_c=\epsilon_{abc}$, where $\epsilon_{012}=1=-\epsilon^{012}$, it is checked that $[\gamma_a,\gamma_b]=2 \epsilon_{ab}^{\ \ \ c}\gamma_c$ so generators $\JJ_a$ form a $so(2,1) \cong su(2)$ algebra,
\begin{equation}
[\JJ_a,\JJ_b]=\epsilon_{ab}^{\ \ \ c}\JJ_c\,.
\end{equation}
The relation to the conventional double index Lorentz generators is given by
\begin{equation}
 \JJ^a = \half \epsilon^a{}_{bc} \JJ^{bc} \,, \quad \JJ_{ab}=-\epsilon_{abc} \JJ^c\,,
\end{equation}
and $\Sigma_{ab}=(1/4)[\gamma_a,\gamma_b]= (1/2)\epsilon_{ab}^{\ \ \ c}\gamma_c$.

For the internal generators we have the $su(2)$ algebra
\begin{equation}
 [\TT_I,\TT_J]=\epsilon_{IJ}^{\ \ \ K}\TT_K\,,
\end{equation}
and they are anti-Hermitian $\TT_I^\dag=-\TT_I$.

Inlcuding the supercharges all the (anti-)commutators close in a $su(2|2)$ superalgebra
\begin{eqnarray}
&[\JJ_a,\overline{\QQ}_\alpha^i]=\frac{1}{2}\overline{\QQ}_\beta^i(\gamma_a)^\beta_{\ \alpha}\,, \quad [\JJ_a,\QQ^\alpha_i]=-\frac{1}{2}(\gamma_a)^\alpha_{\ \beta}\QQ^\beta_i\,,&\\
&[\TT_I,\overline{\QQ}_\alpha^i]=-\frac{i}{2}\overline{\QQ}_\alpha^j (\sigma_I)_j^{\ i}\,, \quad [\TT_I,\QQ^\alpha_i]=\frac{i}{2}(\sigma_I)_i^{\ j}\QQ^\alpha_j\,,&\\
&\{\QQ^\alpha_i,\overline{\QQ}_\beta^j\}=\delta^j_i(\gamma^a)^\alpha_{\ \beta} \JJ_a-i\delta^\alpha_\beta(\sigma^I)_i^{\ j}\TT_I-i\delta^j_i\delta^\alpha_\beta \mathbb{Z}\,.&
\end{eqnarray}
\subsection*{Super-traces:}
The grading operator is given by
\begin{equation}
\mathcal{G}^A_{\ B} = \delta^A_{\alpha}\delta^\alpha_{\ B}-\delta^{A}_{i}\delta^i_{\ B}\,,
\end{equation}
it classifies generators in bosonic $B=\{\JJ_a,\TT_I,\mathbb{Z}\}$ or fermionic $F=\{\QQ^\alpha_i,\overline{\QQ}^i_\alpha\}$, by $[B,\mathcal{G}]=0=\{F,\mathcal{G}\}$, and it squares to one $\mathcal{G}^2=1$. With the grading operator we can define an invariant supertrace
\begin{equation}
\langle G\rangle  \equiv \text{tr}(\mathcal{G} G)=0\,.
\end{equation}
This grants that
\begin{equation}
 \langle B_1 B_2\rangle =\langle B_2 B_1\rangle \,, \quad \langle B F\rangle =\langle F B \rangle \,, \quad \langle F_1 F_2\rangle =-\langle F_2 F_1\rangle \,.
\end{equation}

All the generators $G$ in the representation are supertraceless
\begin{equation}
\langle G\rangle=0\,, \quad G=\{\JJ_a,\TT_I,\mathbb{Z},\QQ^\alpha_i,\overline{\QQ}^i_\alpha\}\,,
\end{equation}
but some quadratic combinations give nontrivial traces
\begin{eqnarray}
&\langle \JJ_a \JJ_b \rangle=\frac{1}{2}\eta_{ab}\,, \qquad \langle \TT_I \TT_J \rangle=\frac{1}{2}\delta_{IJ}\,,&\\
&\langle \QQ^\alpha_i \overline{\QQ}^j_\beta\rangle=-\delta^\alpha_\beta \delta^j_i=-\langle \overline{\QQ}^j_\beta \QQ^\alpha_i\rangle\,.&
\end{eqnarray}
or
\begin{eqnarray}
&\langle \JJ_{ab} \JJ_{cd} \rangle=\frac{1}{2}(\eta_{ac}\eta_{bd}-\eta_{ad}\eta_{bc})\,, \qquad \langle \TT_I \TT_J \rangle=\frac{1}{2}\delta_{IJ}\,,& \label{eq1}\\
&\langle \QQ^\alpha_i \overline{\QQ}^j_\beta\rangle=-\delta^\alpha_\beta \delta^j_i=-\langle \overline{\QQ}^j_\beta \QQ^\alpha_i\rangle\,.&
\end{eqnarray}

Note that in (\ref{eq1}) $\JJ_{ab}$ stands for $\JJ^+_{ab}$ or $\JJ^-_{ab}$ and the traces for
\begin{equation}
 \JJ_{ab} = \JJ^+_{ab} + \JJ^-_{ab}\,,
\end{equation}
is twice as in (\ref{eq1}).

\end{appendices}

\bibliographystyle{ieeetr}
\bibliography{paper.bib}

\begin{thebibliography}{10}

\bibitem{Alvarez:2011gd}
P.~D. Alvarez, M.~Valenzuela, and J.~Zanelli, ``{Supersymmetry of a different
  kind},'' {\em JHEP}, vol.~04, p.~058, 2012.

\bibitem{Alvarez:2015bva}
P.~D. Alvarez, P.~Pais, E.~Rodr\'\i{}guez, P.~Salgado-Rebolledo, and
  J.~Zanelli, ``{Supersymmetric 3D model for gravity with $SU(2)$ gauge
  symmetry, mass generation and effective cosmological constant},'' {\em Class.
  Quant. Grav.}, vol.~32, no.~17, p.~175014, 2015.

\bibitem{Sohnius:1985qm}
M.~F. Sohnius, ``{Introducing Supersymmetry},'' {\em Phys. Rept.}, vol.~128,
  pp.~39--204, 1985.

\bibitem{Sa:1999yf}
P.~M. Sa, ``{Polytropic stars in three-dimensional space-time},'' {\em Phys.
  Lett. B}, vol.~467, p.~40, 1999.

\bibitem{Gundlach:2020ovt}
C.~Gundlach and P.~Bourg, ``{Rigidly rotating perfect fluid stars in $2+1$
  dimensions},'' {\em Phys. Rev. D}, vol.~102, no.~8, p.~084023, 2020.

\bibitem{Murshid:2021iaq}
M.~Murshid, N.~Rahman, I.~Radinschi, and M.~Kalam, ``{Analytical model of low
  mass strange stars in 2+1 spacetime},'' 12 2021.

\bibitem{Eiroa:2020dip}
E.~F. Eiroa and G.~Figueroa-Aguirre, ``{Thin shells in ( 2+1 )-dimensional
  $F(R)$ gravity},'' {\em Phys. Rev. D}, vol.~103, no.~4, p.~044011, 2021.

\bibitem{Banerjee:2014mwa}
A.~Banerjee, F.~Rahaman, K.~Jotania, R.~Sharma, and M.~Rahaman, ``{Exact
  solutions in (2+1)-dimensional anti-de Sitter space-time admitting a linear
  or non-linear equation of state},'' {\em Astrophys. Space Sci.}, vol.~355,
  pp.~353--359, 2015.

\bibitem{Alvarez:2022bqe}
P.~D. Alvarez and J.~Ortiz, ``{Spinor solutions of a Chern\textendash{}Simons
  model for the superconformal algebra},'' {\em Class. Quant. Grav.}, vol.~39,
  no.~24, p.~245007, 2022.

\bibitem{Alvarez:2013tga}
P.~D. Alvarez, P.~Pais, and J.~Zanelli, ``{Unconventional supersymmetry and its
  breaking},'' {\em Phys. Lett. B}, vol.~735, pp.~314--321, 2014.

\bibitem{Alvarez:2020qmy}
P.~D. Alvarez, M.~Valenzuela, and J.~Zanelli, ``{Chiral gauge theory and
  gravity from unconventional supersymmetry},'' {\em JHEP}, vol.~07, no.~07,
  p.~205, 2020.

\bibitem{Alvarez:2020izs}
P.~D. Alvarez, M.~Valenzuela, and J.~Zanelli, ``{Role of gravity in particle
  physics: A unified approach},'' {\em Int. J. Mod. Phys. D}, vol.~29, no.~11,
  p.~2041012, 2020.

\bibitem{Alvarez:2021zhh}
P.~D. Alvarez, L.~Delage, M.~Valenzuela, and J.~Zanelli, ``{Unconventional SUSY
  and Conventional Physics: A Pedagogical Review},'' {\em Symmetry}, vol.~13,
  no.~4, p.~628, 2021.

\bibitem{Alvarez:2021zsw}
P.~D. Alvarez, R.~A. Chavez, and J.~Zanelli, ``{Gauging the superconformal
  group with a graded dual operator},'' {\em JHEP}, vol.~02, p.~111, 2022.

\bibitem{Alvarez:2014uda}
P.~D. Alvarez, P.~Pais, E.~Rodr\'\i{}guez, P.~Salgado-Rebolledo, and
  J.~Zanelli, ``{The BTZ black hole as a Lorentz-flat geometry},'' {\em Phys.
  Lett. B}, vol.~738, pp.~134--135, 2014.

\bibitem{Andrianopoli:2024twc}
L.~Andrianopoli, R.~Noris, M.~Trigiante, and J.~Zanelli, ``{Supersymmetric
  States in Anti\textendash{}de Sitter D=3 Supergravity with Chiral Torsion},''
  {\em Phys. Rev. Lett.}, vol.~133, no.~3, p.~031602, 2024.

\bibitem{Dasgupta:1998jg}
A.~Dasgupta, ``{Emission of fermions from BTZ black holes},'' {\em Phys. Lett.
  B}, vol.~445, pp.~279--286, 1999.

\bibitem{Gentile:2013nha}
L.~G.~C. Gentile, P.~A. Grassi, and A.~Mezzalira, ``{Fermionic Corrections to
  Fluid Dynamics from BTZ Black Hole},'' {\em JHEP}, vol.~11, p.~153, 2015.

\bibitem{Achucarro:1986uwr}
A.~Ach\'ucarro and P.~K. Townsend, ``{A Chern-Simons Action for
  Three-Dimensional anti-De Sitter Supergravity Theories},'' {\em Phys. Lett.
  B}, vol.~180, p.~89, 1986.

\bibitem{MacDowell:1977jt}
S.~W. MacDowell and F.~Mansouri, ``{Unified Geometric Theory of Gravity and
  Supergravity},'' {\em Phys. Rev. Lett.}, vol.~38, p.~739, 1977.
\newblock [Erratum: Phys.Rev.Lett. 38, 1376 (1977)].

\bibitem{Giribet:2024nwg}
G.~Giribet, O.~Mi{\v{s}}kovi{\'c}, N.~Yazbek, and J.~Zanelli, ``{Naked BPS
  singularities in AdS$_{3}$ supergravity},'' {\em J. Phys. A}, vol.~58, no.~2,
  p.~025201, 2025.

\bibitem{Banados:1992wn}
M.~Ba{\~n}ados, C.~Teitelboim, and J.~Zanelli, ``{The Black hole in
  three-dimensional space-time},'' {\em Phys. Rev. Lett.}, vol.~69,
  pp.~1849--1851, 1992.

\bibitem{Andrianopoli:2023dfm}
L.~Andrianopoli, B.~L. Cerchiai, R.~Noris, L.~Ravera, M.~Trigiante, and
  J.~Zanelli, ``{New torsional deformations of locally AdS3 space},'' {\em
  Phys. Rev. D}, vol.~108, no.~4, p.~044011, 2023.

\bibitem{Mielke:1991nn}
E.~W. Mielke and P.~Baekler, ``{Topological gauge model of gravity with
  torsion},'' {\em Phys. Lett. A}, vol.~156, pp.~399--403, 1991.

\bibitem{Blagojevic:2003vn}
M.~Blagojevi{\'c} and M.~Vasili{\'c}, ``{3-D gravity with torsion as a
  Chern-Simons gauge theory},'' {\em Phys. Rev. D}, vol.~68, p.~104023, 2003.

\bibitem{Garcia:2003nm}
A.~A. Garc{\'i}a, F.~W. Hehl, C.~Heinicke, and A.~Mac{\'i}as, ``{Exact vacuum
  solution of a (1+2)-dimensional Poincar{\'e} gauge theory: BTZ solution with
  torsion},'' {\em Phys. Rev. D}, vol.~67, p.~124016, 2003.

\bibitem{Brown:1986nw}
J.~D. Brown and M.~Henneaux, ``{Central Charges in the Canonical Realization of
  Asymptotic Symmetries: An Example from Three-Dimensional Gravity},'' {\em
  Commun. Math. Phys.}, vol.~104, pp.~207--226, 1986.

\bibitem{Miskovic:2006tm}
O.~Mi{\v{s}}kovi{\'c} and R.~Olea, ``{On boundary conditions in
  three-dimensional AdS gravity},'' {\em Phys. Lett. B}, vol.~640,
  pp.~101--107, 2006.

\bibitem{Gentile:2012tu}
L.~G.~C. Gentile, P.~A. Grassi, and A.~Mezzalira, ``{Fermionic Wigs for BTZ
  Black Holes},'' {\em Nucl. Phys. B}, vol.~871, pp.~393--402, 2013.

\bibitem{Briceno:2021dpi}
M.~Brice{\~n}o, C.~Mart{\'\i}nez, and J.~Zanelli, ``{Overspinning naked
  singularities in AdS3 spacetime},'' {\em Phys. Rev. D}, vol.~104, no.~4,
  p.~044023, 2021.

\end{thebibliography}

\end{document}